\documentclass[conference]{IEEEtran}
\IEEEoverridecommandlockouts
\usepackage{cite}
\usepackage{amsmath,amssymb,amsfonts}
\usepackage{graphicx}
\usepackage{textcomp}
\usepackage{xcolor}

\usepackage{graphicx}
\usepackage{multirow}
\usepackage{listings}
\usepackage[caption=false,font=footnotesize]{subfig}
\usepackage{url}
\graphicspath{{pdf/}{jpeg/}{png/}{figures/}}
\usepackage{multirow}
\usepackage{listings}
\usepackage[caption=false,font=footnotesize]{subfig}
\usepackage{relsize}
\usepackage{comment}

\usepackage{stfloats}
\usepackage{balance}
\usepackage{soul}
\usepackage[normalem]{ulem}
\usepackage{cancel}
\usepackage{tabularx}
\usepackage{algorithm}
\usepackage{algorithmicx}
\usepackage[noend]{algpseudocode}
\usepackage{memhfixc}
\usepackage{subfiles}
\usepackage[english]{babel}
\usepackage[utf8]{inputenc}
\usepackage{soul}  
\usepackage{relsize}

\usepackage{dutchcal}

\usepackage[letterpaper, left=0.6in, right=0.6in, bottom=0.95in, top=0.71in]{geometry}

\usepackage{breqn}

\begin{document}

\title{Enabling Opportunistic Low-cost Smart Cities By Using Tactical Edge Node Placement
\thanks{This material is based upon work supported by the National Science Foundation under Grant No. 1952181.}}

\author{
\IEEEauthorblockN{Oluwashina Madamori\IEEEauthorrefmark{1}, Esther Max-Onakpoya\IEEEauthorrefmark{1}, Gregory D. Erhardt\IEEEauthorrefmark{2}, Corey E. Baker\IEEEauthorrefmark{1}}
    \IEEEauthorblockA{
        \IEEEauthorrefmark{1}Department of Computer Science, \IEEEauthorrefmark{2}Department of Civil Engineering, 
        University of Kentucky, Lexington, KY USA
    }
    \IEEEauthorblockA{
        shina@uky.edu, esther.max05@uky.edu, greg.erhardt@uky.edu, baker@cs.uky.edu 
    }
}

\maketitle

\begin{abstract}
Smart city projects aim to enhance the management of city infrastructure by enabling government entities to monitor, control and maintain infrastructure efficiently through the deployment of Internet-of-things (IoT) devices.
However, the financial burden associated with smart city projects is a detriment to prospective smart cities. 
A noteworthy factor that impacts the cost and sustainability of smart city projects is providing cellular Internet connectivity to IoT devices.
In response to this problem, this paper explores the use of public transportation network nodes and mules, such as bus-stops as buses, to facilitate connectivity via device-to-device communication in order to reduce cellular connectivity costs within a smart city.
The data mules convey non-urgent data from IoT devices to edge computing hardware, where data can be processed or sent to the cloud.
Consequently, this paper focuses on edge node placement in smart cities that opportunistically leverage public  transit networks for reducing reliance on and thus costs of cellular connectivity.
We introduce an algorithm that selects a set of edge nodes that provides maximal sensor coverage and explore another that selects a set of edge nodes that provide minimal delivery delay within a budget. 
The algorithms are evaluated for two public transit network data-sets: Chapel Hill, North Carolina and Louisville, Kentucky. 
Results show that our algorithms consistently outperform edge node placement strategies that rely on traditional centrality metrics (betweenness and in-degree centrality) by over 77\% reduction in coverage budget and over 20 minutes reduction in latency.
\end{abstract}


\section{Introduction}
Development of smarter cities has been proposed as a means of combating the challenges arising from the increasing rate of urbanization within many cities in the world~\cite{Hayat2016}. 
One major characteristic of most smart city designs is the deployment of a vast number of IoT devices/sensors across the city to monitor and sometimes control the state of public infrastructure such as water and gas pipes~\cite{guevara2020role}. 
These IoT devices, which include weather sensors, traffic monitors, parking meters/monitors, generate large amounts of data that need to be forwarded to the Cloud for processing and storage.
To achieve this, deployed IoT devices typically rely on cellular connectivity. 
However, the additive operating costs incurred from each sensor's cellular subscription plans is expensive~\cite{paradells2014}. 
For example, cities such as San Diego, New Orleans, London, and Songdo have either proposed or invested in smart city projects that cost between \$30 Million and \$40 Billion.
The costs of deploying and maintaining smart city projects is a huge deterrent for city officials, especially when the sustainability and impact of such projects are uncertain~\cite{Deloitte2018,10.1145/3233985,SmartCitiesWorld}. 
In addition, since sensors generate large amounts of data and connect to the same base stations that facilitate cellular connectivity for personal mobile devices, solely using cellular networks for smart city data can quickly lead to network congestion and poor user experience. Though 5G has been proposed as a viable solution, projections show that 5G will not be able to support the load of billions of IoT devices coming online~\cite{cisco,loghin2020disruptions,navarro2020survey}. 
Hence, there is need for cost-effective smart city communication networks that reliably and efficiently forward sensor data to the cloud without over burdening cellular infrastructure. 
In response to aforementioned needs, various researchers have historically explored delay tolerant networks (DTNs) or opportunistic networks for smart city applications that can tolerate high latency~\cite{Choi2016, giannini2016, Madamori2019, amah2020preparing}.

Opportunistic networks are attractive because of their ability to persist data with minimal infrastructure.
Such networks leverage the already existing mobility of nodes within a city to retrieve data from IoT  devices and either disseminate data to other  devices  in the network or act as intermediate data carriers which forward the data to specific locations that have edge computation, cloud connectivity, and storage resources~\cite{Choi2016,ho2019next}. 
The research community has investigated the use of public transit vehicles --- such as buses, trams, and light rail --- for opportunistic networks; however most of the research has been limited to routing and forwarding schemes~\cite{baker2017vivo}.
Messages are delivered with some delay which is directly correlated with the layout, density, and mobility of nodes in the network~\cite{hui2008phase,Keraenen2009}.
Consequently, a question that has been marginally addressed is: \textbf{where should edge nodes be placed in a low-cost smart city that leverages public transit networks to improve connectivity?}

The opportunistic use of transportation networks is not intended to perform better than 5G or other centralized communication schemes, but instead offer a low-cost alternative for delivering time-insensitive data, enabling municipalities to become smart cities at a fraction of the cost. In this paper we:
(i) introduce the Maximal Sensor Coverage (MSC) edge node placement optimization problem, and explore the Minimal Delivery Delay (MDD) problem;
(ii) formulate the Maximal Sensor Coverage (MSC) as a set cover problem;
(iii) develop approximation algorithms for solving the optimization problems highlighted; and finally;
(iv) compare the results of our algorithms with traditional network centrality measures.

The rest of the paper is structured thus: 
Section \ref{sec:rev_relatedworks} discusses related work;
Section \ref{sec:model} introduces the network model;
Section \ref{sec:opt} defines the MDD problem and describes its solution;
Section \ref{sec:sim} explains the simulation design and environment;
Section \ref{sec:eval} offers the numerical evaluation; and finally
Section \ref{sec:discussion} discusses and concludes the work.

\section{Related Works} \label{sec:rev_relatedworks}
Various researchers have investigated the opportunistic use of vehicular transportation networks for data forwarding in smart city communication networks.
The network architecture often consists of a set of vehicular data mules (e.g. buses, boats, train, etc) that encounter IoT devices, opportunistically collect sensed data and deliver the data to edge nodes with wired Internet connectivity~\cite{almeida2018multi,raissi2019autonomous}.
However, the vast majority of research in this area have focused on the design of data forwarding schemes.
Some works have proposed new routing schemes that harness the quasi-deterministic nature of public transportation networks and utilize metrics such as intercontact times to reduce latency and improve delivery~\cite{Choi2016,giannini2016}.
Others have explored routing in the context of different wireless communication media such as, LoRa and WiFi~\cite{almeida2018multi}.
Some others have investigated routing and forwarding schemes in the context of preserving privacy~\cite{amah2020preparing}.
Nevertheless, none of the aforementioned works addresses the question of edge node placement.

While placement algorithms are typically unique to the network, certain techniques from other network domains are relevant to this problem.
Some previous works on placement optimization for networks similar to ours include~\cite{Krause2008EfficientNetworks} which exploits the principle of submodularity to tackle the problem of sensor placements in water distribution networks.
In~\cite{xhafa2015solving}, the authors explore several optimization techniques for access point placement in wireless mesh networks. 
Additionally,~\cite{bagula2015design} looked at optimizing coverage and cost-efficiency in a smart parking network.
Unlike prior research, this work proposes an edge node placement algorithm for low-cost smart cities that leverages opportunistic networks and evaluates it using GTFS-data derived from real transits from multiple cities.

\subsection{Classic Centrality Analysis}
The problem of edge node placement is similar to finding the most significant stops in a transit network. 
In the field of network analysis, there exist several popular centrality measures. 
These centrality measures are usually computed by a real-valued function and reflects a node's significance or importance within the respective network~\cite{Newman:2010:NI:1809753}. 
Centrality measures have been used in many kinds of networks including the Internet, social networks, biological networks, and transportation networks. 
Unfortunately, centrality measures work best with simple static networks~\cite{chen2017interplay} and not dynamic networks. 
Since our network model is more complex, containing not just nodes (stops) and edge nodes (trips), but also vehicle schedule information across each node, centrality measures may not provide the best solutions for optimizing edge node placement.
Hence, for the optimization problems considered in this work, we explore other solutions. 

\subsection{Evaluation of Opportunistic Networks}
A number of test-beds and simulation tools have been developed by the research community and presented in literature to facilitate research in the field of vehicular communication networks and vechicular opportunistic networks.
In~\cite{Soroush2009}, the authors create the DOME testbed, to give researchers access to transit buses already furnished with necessary equipment, so that external researchers can upload their communication protocols to the buses remotely and conduct experiments within a real-world environment without having to invest in additional infrastructure.
In \cite{paradells2014}, the authors also design an ad-hoc 
testbed using buses. 
Although real-world deployments of various network architectures provide the most accurate results, oftentimes these results are not sufficiently generalized due to insufficient geographic diversity as well as the limited scale of the experiments. In addition, real-world deployments are typically financially expensive.

One popular alternative to real-world evaluations has been the use of mathematical models and simulations.
Some of the advantages of this approach include: minimal financial costs, the ability to easily scale up experiments, and the heterogeneity of geography and hardware parameters during experiments. 
Examples of some widely used open-source generic simulators in the field of vehicular communication and DTN include the ONE simulator~\cite{Keraenen2009} and ns-3~\cite{Henderson2008NetworkSimulator}.
However, it can be difficult to fully incorporate real-world data into simulation environments, or adapt the environment to match the designed network model. Hence, several works in literature use custom-built simulators for their models.

In \cite{barba2012smart}, the authors designed a simulation environment for VANETS
and intelligent traffic lights to notify vehicles of traffic and warning messages using Ad-hoc On-Demand Distance Vector (AODV).
In \cite{bonola2016opportunistic}, the authors conduct a feasibility study
by setting up a simulation framework that relies purely on opportunistic 
interactions between taxi cabs.  
The lower accuracy of simulation environments, when compared to real-world environments is a factor that often undermines the integrity and validity of results.
An additional contribution of our work is the development a simulator for our network model that closely emulates the movement of real world transit vehicles, hence we are able evaluate the performance of our network across many cities, while also maintaining high result accuracy.


\begin{figure*}[htbp]
    \centering
    \includegraphics[width=0.8\textwidth,height=3.5in]{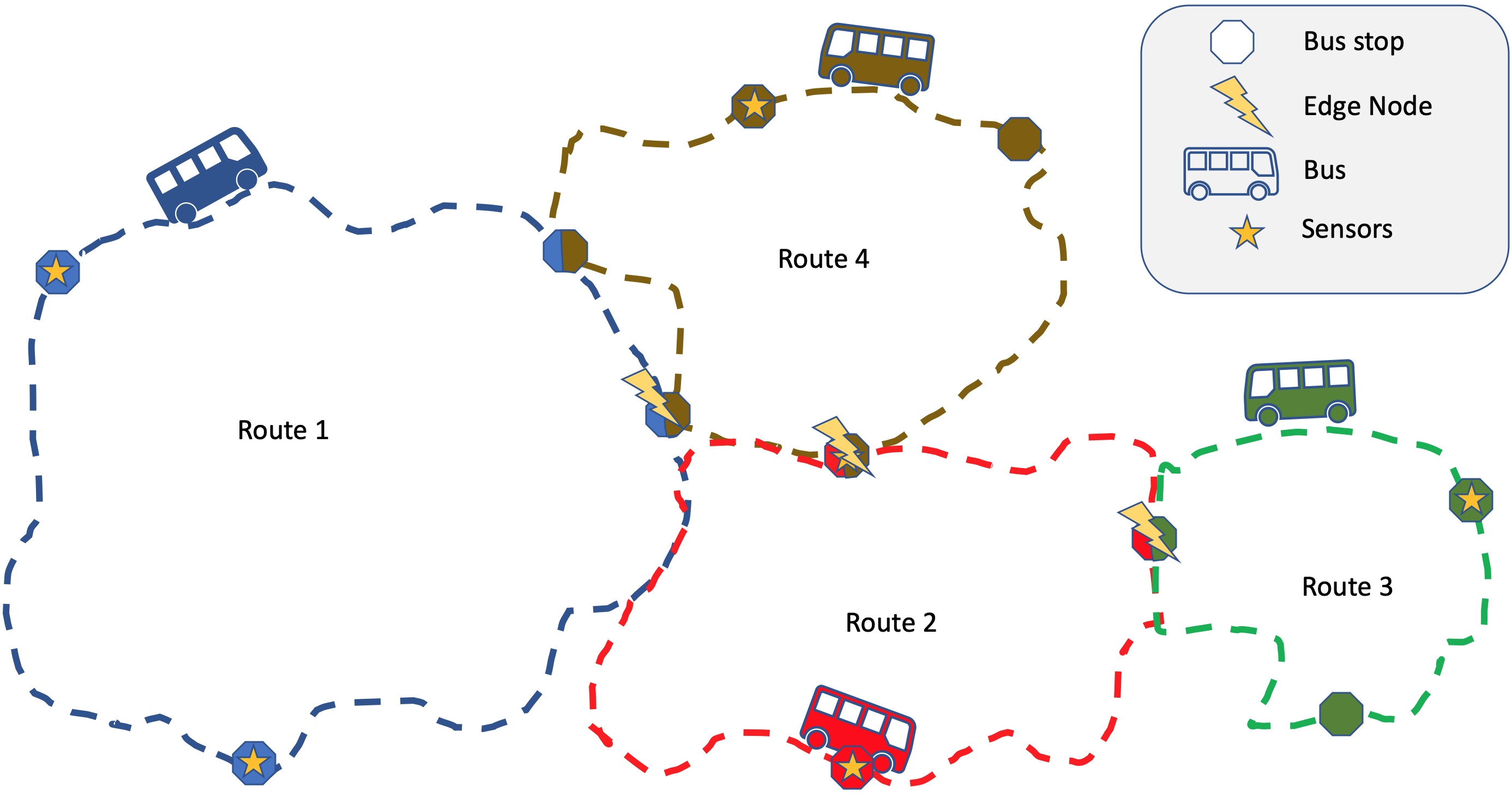}
    \caption{Network Architecture}
    \label{fig:architechture}
\end{figure*}

\section{Model}\label{sec:model}
The entities within the smart city are depicted in Figure~\ref{fig:architechture} and described below

\subsubsection{Bus routes} 
Every public transit network has a list of predefined routes on which buses move/operate. 
Every route is primarily defined by the list of stops on which the route passes through. 
A route also contains a list of trips which specifies the arrival and departure times for buses operating on that route, as well as the sequence of stops the bus moves through during each trip. 
Note that a stop may service more than one route.

\subsubsection{Sensors} 
Sensors are located at select bus stops. 
Each sensor generates data at a specific periodic rate around-the-clock. 
The data is stored locally until it can be forwarded to a bus. 
Every sensor is equipped with a device that allows it to opportunistically connect to a bus when the bus is within a specified geographic range of the sensor. 
In addition, we are only considering sensors whose generated data packets are small enough to justify an assumption of transmitting to a bus in a short duration.
Local storage of a sensor is considered to be infinite, since data is expected to be picked up by a bus within a couple of days, thereby eliminating the need for a policy for dropping packets.
Also, it is assumed that the data generated by these sensors is delay tolerant and is not relied upon to make real-time decisions.

\subsubsection{Edge nodes} 
These are stationary and when ``active (on)" are considered always-connected devices that forward data directly to remote servers via the Internet for post processing and analysis. 
They act as the destination for all data generated by sensors. 
Not all bus stops are edge nodes, rather edge nodes are placed at selected bus stops. 
Similar to sensors, edge nodes are assumed to always be within transmission range of buses traveling on routes for the stop on which the edge node is located. 
Each edge node is also equipped with a device that allows it to opportunistically connect to a bus when the bus is within a specified geographic range of it. 
Edge nodes are equipped with edge computing and pre-processing capabilities.

\subsubsection{Buses} 
These move along predefined routes on a fixed schedule. 
Hence, the specific geographic position of buses moving on a specific route at any time can be estimated. 
Each bus is equipped with the necessary hardware to connect to sensors, retrieve, store sensor data packets, and forward the data to edge nodes. 
We currently do not consider the buffer/queue sizes of data in buses and so buses are assumed to have buffer/queue sizes of infinity since they are expected to deliver data to at least one gateway within a day (buses park at transit stations which are expected to have an active gateway), so a drop-policy for stored data is not necessary.


\section{Edge node Selection Problem}\label{sec:opt}
In this section, the edge node selection problem is represented as two optimization objectives and corresponding algorithmic solutions are provided. 
It is assumed that the cost associated with installing or activating a gateway is constant regardless of the location being considered.

\subsection{Maximal Sensor Coverage (MSC)}\label{subsec:coverage}
The objective, Maximizing Sensor Coverage (MSC), is to \textit{find the minimum/smallest set of locations to install or activate gateways such that there is at least one direct path in the network from all possible sensor locations to an edge node}.
This would ensure that all sensors, regardless of where they are placed on the transit network will have a chance of having its data delivered.

Recall that the bus transit network consists of several routes through which vehicles move, each route comprises of a set of stops and two or more routes may share a stop. 
Therefore, we can define our MSC problem as such: \textit{Given a set of routes $R$, and a list of stops, $S$, each with a subset of routes, $R
_l \subseteq R$, that use that stop (each element of $R$ is associated with at least one stop), find the minimal set of stops required to cover all routes}. 
This problem can be reduced to a minimal set cover problem. 
In the set cover problem, we are given a universal set U, such that $|U|= n$, and a family of subsets $L_1,...,L_k  \subseteq U$. A set cover is a collection C of the subsets $L_1,...,L_k$  whose union is the universal set $U$. Formally, C is a set cover if $\bigcup_{L_i \in C} L_i = U$. To find the minimal set cover, the objective is to minimize $|C|$.

The reduction is fairly intuitive.
In our case the universal set is the set of all routes $R = U$, and the family of subsets are the set of routes each stop services, $S = L$.
Given a decision variable, $x_l \in \{0, 1\}$, which indicates whether a stop in $L$ is picked, the ILP formulation is thus:
\begin{equation}\label{}
\text{minimize } \sum_{l \in L} x_l \text{ st}\
\end{equation}
\begin{equation}\label{}
\sum_{l:r \in L} x_l \geq 1 \ \ \ \ \forall \ r \in R\
\end{equation}
\begin{equation}\label{}
x_l =\{0, 1\} \ \ \ \ \forall \ l \in L\
\end{equation}

The problem of finding the optimal set cover solution is \textit{NP-Hard}. 
Nevertheless, the greedy approach is able to find a solution close to the optimal set cover.
It is bounded above by a $O(log_en)$ approximation to optimal solution of the set cover problem, where $n$ is the number of routes in the network.
The greedy MSC algorithm is described in Algorithm \ref{alg:greedy_msc}. 
At each iteration, we find the gateway candidate that provides the largest  increase in the number of routes covered, and add it to the gateway set. 
This process is repeated until all routes have been covered by the gateway set.

\begin{algorithm}
\caption{Maximal Sensor Coverage (MSC)}\label{alg:greedy_msc}
\hspace*{\algorithmicindent} \textbf{Input} — routes \textbf{R}, route subsets \textbf{S} \\
\hspace*{\algorithmicindent} \textbf{Output} — selected gateway set \textbf{G}
\begin{algorithmic}[1]
\Procedure{GREEDY-MSC}{$R,S$}
\State $X \gets R$
\State $G \gets \emptyset $
\While{$X \not=0$} 
\State \texttt{Select an $S_i \subseteq R$ that maximizes} $|S_i \cap X|$
\State $X \gets X \setminus S_i$
\State $G \gets G \cup {S_i}$
\EndWhile\label{euclidendwhile1}
\State \textbf{return} $G$ 
\EndProcedure
\end{algorithmic}
\end{algorithm}

\subsection{Minimal Delivery Delay (MDD)}
The objective of MDD \textit{is to select the set of locations, $G \subset S$, to place edge nodes such that the average network latency for data generated across the network is minimized, without exceeding a budget constraint, $k$}. 
Where the budget constraint refers to the number of edge nodes that can be added to the network and $S$ is the set of stops. 
In considering network latency, we account for both delivered and undelivered data. 
We also assign a penalty value to data that is undelivered.
The penalty value is selected in such a way that it indicates that the message was not delivered within the time window of the simulation.
In our evaluation, we picked a fixed value outside the range of simulation window.
Further, the edge node selection process is conducted without prior knowledge of the stops in which the sensors will placed in the city. 

The problem is formulated as an Influence Maximization (IM) problem, which is defined: Given a network with $n$ nodes and given a propagation process on that network, choose a set of nodes called the \textit{seed set} $D$ of size $b<n$ that maximizes the number of nodes in the network that are ultimately influenced~\cite{Kempe2003MaximizingNetwork}. 
However, our problem differs from the traditional IM problem because the set of nodes we want to select are not seed/source nodes, but destination nodes. 
Hence, we consider each potential edge node location, $s \in S$, to possess an influence value, $\sigma(s)$, which describes its impact on the network latency if it is added to an existing set of edge nodes.
Thus, our problem objective is to select a set of edge nodes, $G$, below a specified cardinality, $k < |S|$, that together decreases the network latency the most. 
Since, $G = D$, $k = b$ and $|S| = n$, the problem can be defined as an influence maximization problem.

Current IM algorithms require an influence function that simulates the propagation process and computes the marginal influence that each potential seed has on the overall propagation. 
Therefore for our algorithm, we develop an influence function ($\sigma$) that computes the expected latency across the network for any potential set of edge nodes.
Given an undelivered messages time penalty - $T$, the minimum time it takes to get from stop, $v$, to stop, $u$ - $\tau(u,v)$, and the lag between the time at which data is generated and the time at which the next bus for the route arrives at the stop where the sensor, $\mathbcal{s}$, is located - $\mathbcal{t}(s, \mathbcal{s})$, the influence function for a set of edge nodes, $G$ is defined as:
\begin{equation}\label{eqn:inflmaxgate}
    \sigma(G) = T - \frac{1}{|S|-G}\sum_{\mathbcal{s} \in S, \mathbcal{s} \not\in G} min(\mathcal{T}(G, \mathbcal{s}))
\end{equation}
Where,
\begin{equation}\label{eqn:inflmaxgateext}
    \mathcal{T}(G, \mathbcal{s}) = \{(\tau(g, \mathbcal{s}) + \mathbcal{t}(g, \mathbcal{s})) \ | \ \forall \ g \in G \}
\end{equation}

Each element in the set, $\mathbcal{T}(G,S)$, is the sum of the time it takes for a vehicle to forward data to an edge node and the time it takes for the vehicle to get to the sensor.
The influence function makes use of ``data delivery delay'' (Algorithm \ref{alg:delay}), to calculate the network latency. 
This function is submodular and its proof can found in a previous work ~\cite{madamori2021percom}.
Even though finding the set of edge nodes that maximize influence is \textit{NP-Hard}, since the influence function is submodular, the solution can be approximated using the \textit{Greedy} and \textit{Cost-Effective Lazy Forward} (CELF) algorithms~\cite{Nemhauser2014}.

\subsubsection{Greedy Algorithm}\label{subsec:greedy_im}
Since the problem is reduced to the maximization of a monotone submodular function, the greedy algorithm provides a $(1-1/e)$ - approximation~\cite{Nemhauser2014}. 
Hence, the greedy algorithm is theoretically guaranteed to choose a gateway set whose network latency will be at least 63\% of the network latency of the optimal gateway set. 
Our greedy algorithm is described in Algorithm~\ref{alg:greedy_im}. 
It starts with an empty gateway set $S = \emptyset$. 
In each iteration, the greedy heuristic chooses a new gateway $u$ from the non-gateway nodes $V \setminus S$ with largest (marginal) influence gain $\sigma(S \cup {u}) - \sigma(S)$ and adds $u$ to $S$. 
The algorithm terminates after selecting $k$ gateways. 

\begin{algorithm}
\caption{Greedy Minimal Delivery Delay}\label{alg:greedy_im}
\hspace*{\algorithmicindent} \textbf{Input} — network graph \textbf{N}, influence function $\sigma$, budget \textbf{k} \\
\hspace*{\algorithmicindent} \textbf{Output} — selected gateway set \textbf{G}
\begin{algorithmic}[1]
\Procedure{GREEDY-MDD}{$N, \sigma, k$} 
\State $G \gets \emptyset $
\While{$|G| < k $} 
\State $ u \gets arg\; max_{\;v \in V \setminus G} \; {\sigma(G \cup {v}) - \sigma(G)}$
\State $G \gets G \cup \{u\}$
\EndWhile\label{euclidendwhile2}
\State \textbf{return} $G$
\EndProcedure
\end{algorithmic}
\end{algorithm}

\subsubsection{CELF-MDD Algorithm}\label{subsec:celf_im}
Although the greedy algorithm is much quicker than a brute-force approach, the greedy algorithm is still very slow when considering the size of actual transit networks. 
Therefore, we use the cost-effective lazy forward (CELF) approach.
CELF significantly reduces the running time by exploiting the submodular property of our influence function while still providing the same solution set as the Greedy algorithm~\cite{Leskovec2007Cost-effectiveNetworks}.
It eliminates the need to compute the marginal influence value of all potential edge nodes at each iteration. 

In the first round, we calculate the influence for all stops (like Greedy), select the stop with the greatest influence, and store the influence values of the other stops in a max heap.
In subsequent iterations, the marginal influence of the top stop in the heap is computed and added back to the heap. 
If the stop remains at the top of the heap, then it must have the highest marginal influence of all remaining stops, due to the submodular property of the influence function. 
If a different stop is on top of the heap, the process continues until a stop remains on top after two iterations, after which that stop is added to the edge node set. 
This process is repeated until the edge node budget has been met.

\begin{algorithm}
\caption{CELF Minimal Delivery Delay (CELF-MDD)}\label{alg:celf_im}
\hspace*{\algorithmicindent} \textbf{Input} — graph \textbf{$N$}, influence function \textbf{$\sigma$}, budget \textbf{$k$} \\
\hspace*{\algorithmicindent} \textbf{Output} — selected edge node set \textbf{$G$}
\begin{algorithmic}[1]
\Procedure{CELF-MDD}{$N, \sigma, k$} 
\State $G \gets \emptyset $
\State $Q \gets \emptyset $
\For{$v \in N.nodes$}
\State $u \gets v$
\State $u.gain = \sigma (\{v\})$
\State \texttt{add $u$ to $Q$ in descending order}
\EndFor
\While{$|G| < k $}
\State $u \gets Q.top$
\If{$u.flag = |G|$}
\State $G \gets G \cup \{u\}$
\State $Q \gets Q \setminus {u}$
\Else
\State $u.gain \gets \sigma(G \cup \{u\}) - \sigma(G)$
\State $u.flag \gets |G|$
\State \texttt{Re-sort $Q$ in descending order}
\EndIf
\EndWhile\label{euclidendwhile3}
\State \textbf{return} $G$
\EndProcedure
\end{algorithmic}
\end{algorithm}

\section{Simulation Design}\label{sec:sim}
We developed a simulation tool\footnote{The code for the simulator is available on GitHub - \url{https://github.com/netreconlab/low_cost_smart_city_optimization}}
that models a vehicular communication network consisting of sensors, buses and edge nodes within any real-world city, by directly using real transit network information provided in the General Transit Feed Specification (GTFS) format \cite{GTFSDevelopers}.
GTFS handles information on transit routes, stops, and timetables~\cite{GTFSDevelopers}. 
By building a simulation tool that incorporates GTFS information, we are able to evaluate the performance of our low-cost smart city model for hundreds of cities around the world.

\subsubsection{Transit feed to graph conversion} 
The GTFS transit feed data for a transit agency is converted into a directed graph. The conversion is done using an open-source library called \textit{peartree}~\cite{Butts2018Peartree:Analysis.}. 
The graph contains:
(i) Nodes representing stops, with each node containing the departure times for all vehicles from that stop,
and 
(ii) Edges representing a bus path from one stop to another. 
The weight on each edge is the average time it takes for a bus to get from one stop to a neighboring stop on a trip. 

\subsubsection{Sensor Placement} 
Sensors are placed at randomly selected stops in the transit network. 
Each stop has a maximum of one sensor and the total number of sensors to be placed in the network is defined for each simulation.
In addition, each sensor is assigned a time value representing the frequency at which it generates data. 
This frequency value is assigned to each sensor based on a uniform random distribution.

\subsubsection{Data Delivery Delay} \label{subsub:data_delivery}
For each data packet generated at a sensor, the shortest duration it takes for the data packet to get to an edge node is computed.  
First, a subgraph (consisting of only stops in a single route) is extracted from the main graph for each route. 
Next, we iterate through each route the sensor is on and compute the shortest path length from the sensor to any edge node on that route. 
Since the edge node weight is the average travel time for vehicles between a node pair, the computed shortest path length represents the estimated time it takes for a vehicle to forward the data to an edge node after departing from the sensor. 
The total estimated delay is calculated by adding the shortest path length to the waiting time (the lag between the data generation time and the time at which the next vehicle associated with that route arrives at the stop where the sensor is located).

After iterating through all routes for the sensor, the path with the shortest total estimated delay is designated as the path through which the data will travel.
The path length of the designated path is the estimated end-to-end delay for the data packet generated. 
Algorithm~\ref{alg:delay} presents the pseudocode for computing the delay for each data packet generated.

\begin{algorithm}
\caption{Pseudocode for Computing data delivery delay}\label{alg:delay}
\hspace*{\algorithmicindent} \textbf{Input} — routes, $R$; sensor, $E$; time, $t$ \\
\hspace*{\algorithmicindent} \textbf{Output}  — delay, $D$
\begin{algorithmic}[1]
\Procedure{ComputeDelay}{$R, E, t$}
\State $D \gets \infty $
\For{$r \in R$}
\State $G_r \gets \textsc{GetRouteSubgraph}(r)$
\For{$g \in r.gateways$}
\State $length$ $\gets \textsc{DijkstraShortestPath}(G_r, E, g)$
\If{$length$}
\State $wait$ $\gets \textsc{TimeTillNextArrival}(E, R, t)$ 
\If{$wait$}
\State $D \gets min(D, length + wait)$
\EndIf
\EndIf
\EndFor
\EndFor
\State \textbf{return} $D$
\EndProcedure
\end{algorithmic}
\end{algorithm}

\begin{figure*}[htbp]
\centering

    \subfloat[]{\includegraphics[width=0.49\textwidth,height=3.0in]{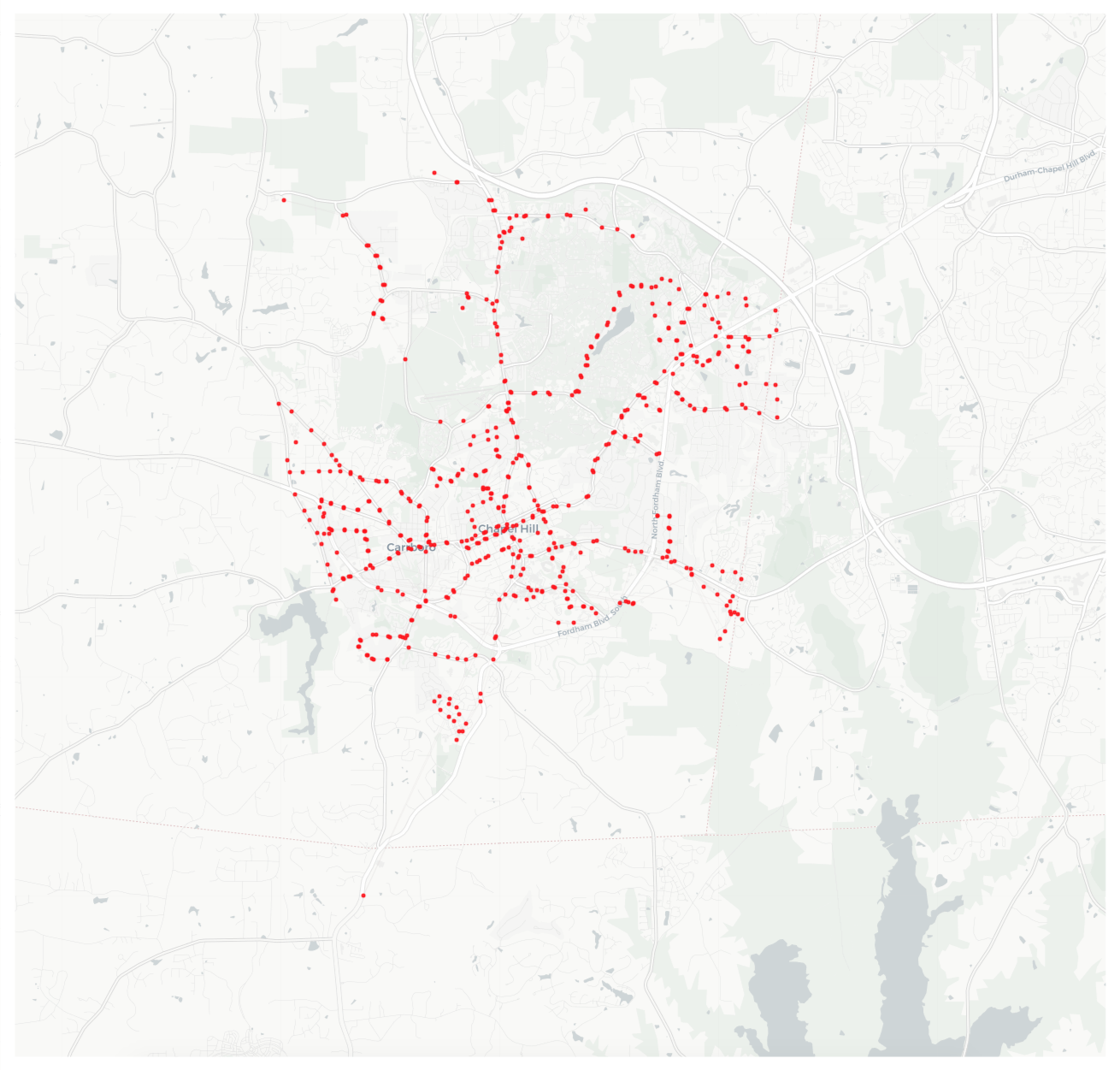}\label{fig:maps}}
	\subfloat[]{\includegraphics[width=0.49\textwidth,height=3.0in]{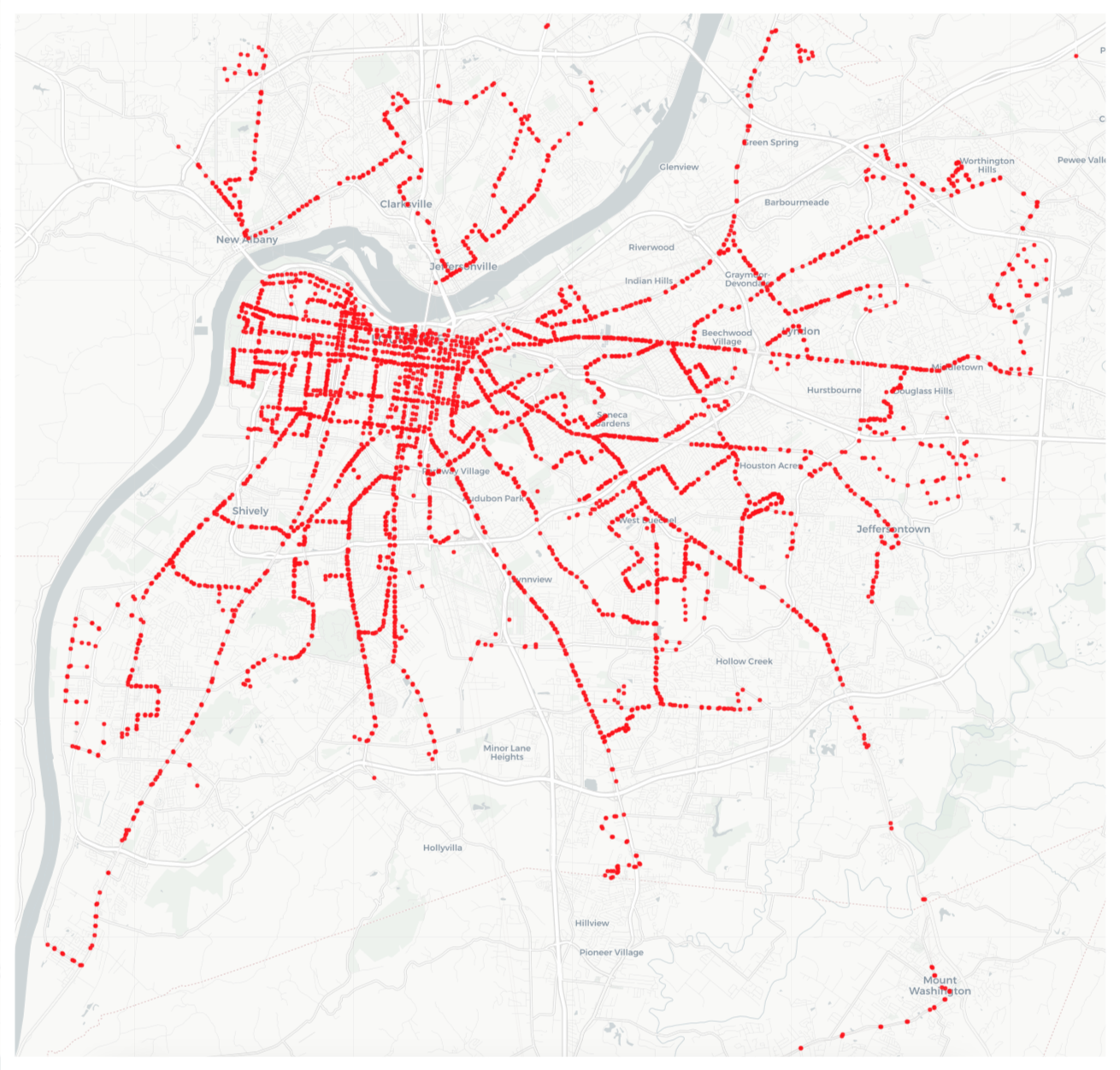}\label{louisville_map}}

\caption{(a) Map of CHT bus-stop locations in Chapel Hill, (b) Map of TARC bus-stop locations in Louisville.} 
\label{results}
\end{figure*}

\subsubsection{Storing results for analysis} 
For each simulation, important information such as generation time, delivery time, delivery path, vehicle wait time, and vehicle travel time for each data packet generated during simulation is recorded and stored in JavaScript Object Notation (JSON) file. 
This helps with carrying out post-simulation analysis after the simulation has ended without having to re-run simulations. 

\section{Numerical Evaluation}\label{sec:eval}
\subsection{Simulation Setup}
In order to evaluate the performance of our algorithm, we make use of the GTFS data of two bus transit agencies; Chapel Hill Transit (CHT) in Chapel Hill, North Carolina, and Transit Authority of River City (TARC) in Louisville, Kentucky~\cite{OpenMobilityOrgOpenMobilityDataWorld}.
Chapel Hill is a relatively small town, measuring 55 km sq (21.3 sq miles) and an estimated population of about 60,988. 
Louisville, on the other hand, is a much larger city with a population of 620,118 and land area of 171.70 km sq (66.29 sq miles)~\cite{CenterforDiseaseControl2016}. 
The difference in city size also translates to the differences in public transit networks present in both cities as highlighted in Table~\ref{tab:network_stats} and Figure~\ref{results}.

\begin{table}[htbp]
\centering
\caption{Bus Network Characteristics CHT and TARC}
\label{tab:network_stats}
\begin{tabular}{|l|l|l|}
\hline
\textbf{Statatistics} & \textbf{CHT} & \textbf{TARC} \\ \hline
Routes        &           26                   &         46                    \\
Stops         &            571                  &       4391                      \\
Total trips   &            1252                  &       1917                      \\ 
Betweenness centrality avg. &    0.04896       &       0.00829                      \\
In-degree centrality avg. &   0.00210         &    0.00025                       \\\hline
\end{tabular}
\end{table}

\subsection{Maximizing Coverage}

\begin{figure*}[htbp]
\centering

    \subfloat[]{\includegraphics[width=0.33\textwidth]{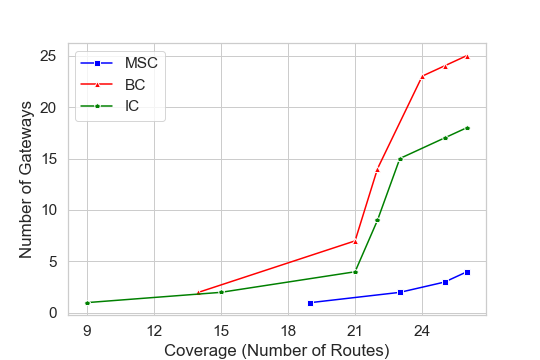}\label{fig:coverage_cht}}
    \subfloat[]{\includegraphics[width=0.33\textwidth]{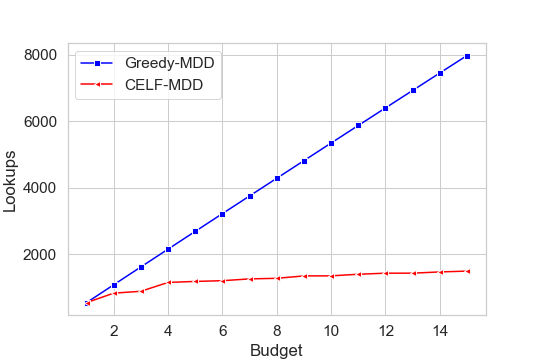}\label{fig:lookups}}
    \subfloat[]{\includegraphics[width=0.33\textwidth]{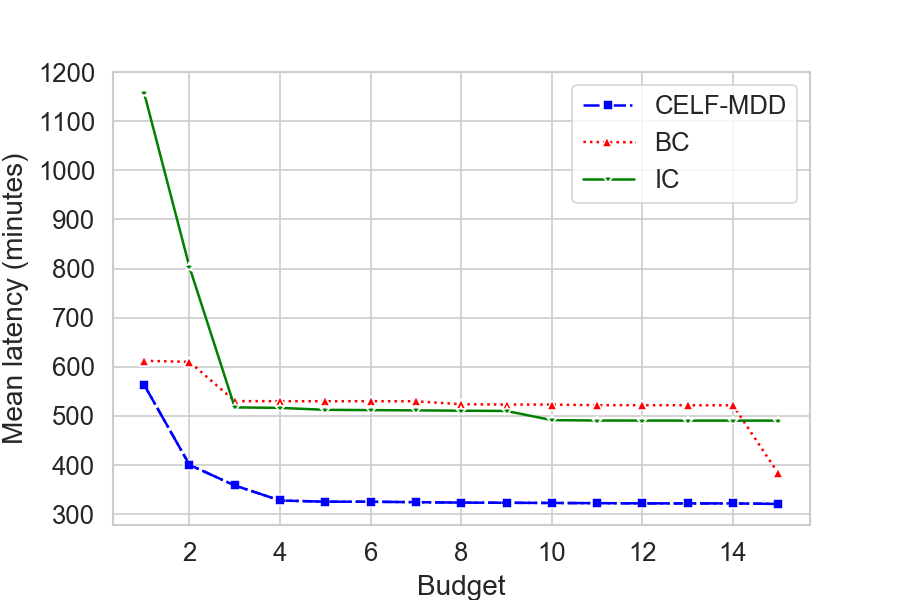}\label{budget_delay_cht}}
    
	\subfloat[]{\includegraphics[width=0.33\textwidth]{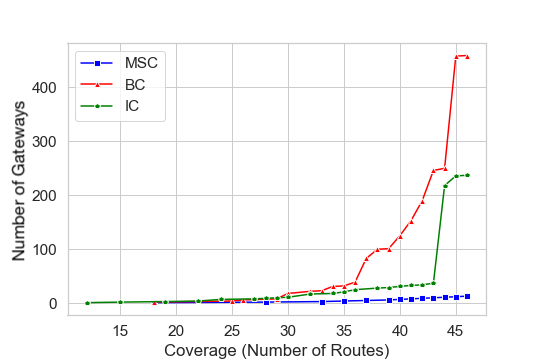}\label{fig:coverage_louisville}} 
	\subfloat[]{\includegraphics[width=0.33\textwidth]{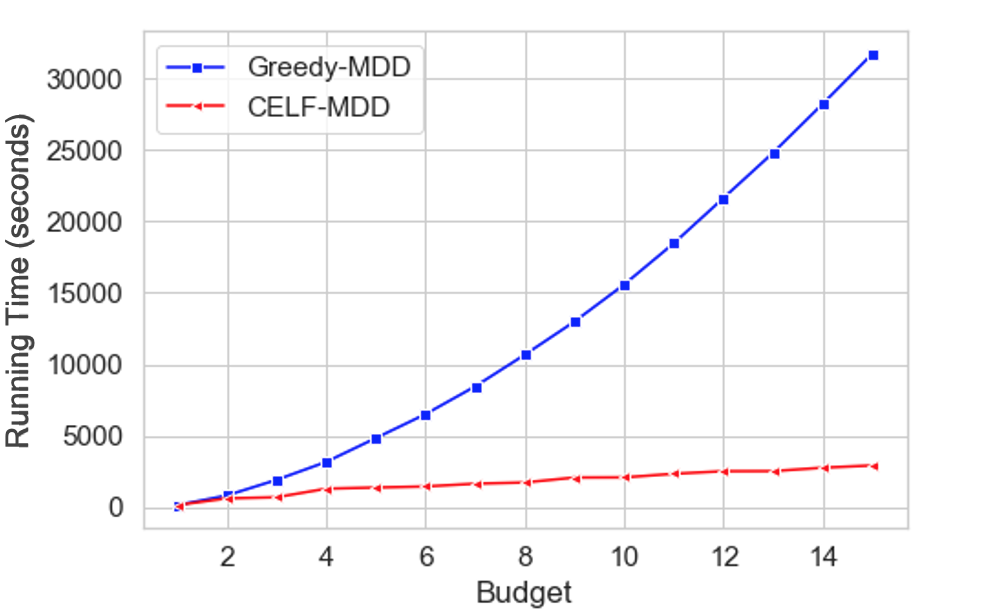}\label{fig:running_time}} 
    \subfloat[]{\includegraphics[width=0.33\textwidth]{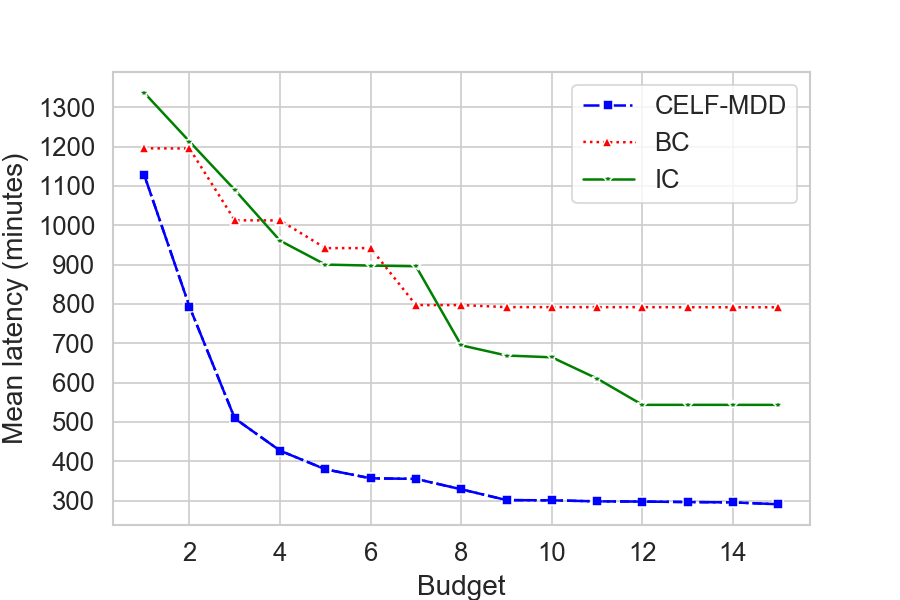}\label{budget_delay_louisville}}

\caption{(a)  MSC performance for CHT, (b) Graph  showing  the  number  of  lookups  for  Greedy-MDD  and CELF-MDD, (c) Average network latency for CHT  (d) MSC performance for TARC. (e) Graph showing the number of running time for Greedy-MDD and CELF-MDD (f) Average network latency for TARC.} 
\label{fig:im_results}
\end{figure*}

Figures~\ref{fig:coverage_cht} and~\ref{fig:coverage_louisville} along with Table \ref{tab:coverage} give insight into the performance of the MSC Algorithm described in \ref{subsec:coverage} when compared two traditional graph centrality measures - betweenness centrality (BC), and in-degree centrality (IC). 
For the centrality measures, we greedily picked each gateway in order of decreasing centrality, until all routes were covered. 
Figures~\ref{fig:coverage_cht} and~\ref{fig:coverage_louisville} show the rate of increase in route coverage compared to the number of gateways selected. 
We see that MSC outperforms BC and IC significantly for both CHT and TARC. 
For CHT, MSC requires just 4 gateways to cover all routes in the network, compared to 18 and 25 gateways in IC and BC, respectively. 
For a larger network like TARC, MSC requires 13 gateways to cover all routes in the network, compared to 257 and 459 gateways in IC and BC, respectively.

Table \ref{tab:coverage} also highlights the delivery ratio for each algorithm. Since the three algorithms (MSC, BC, and IC) cover all routes, the delivery ratio is essentially the same. 
Delivery ratio is less than 100\% because delivery also depends on a bus arriving at a sensor within the simulation time window. 
This means that for some data packets generated, especially in the later stages of the simulation, there were no buses moving on that route causing the data packets to remain undelivered. 

\begin{table}[htbp]
\caption{Results of MSC, BC and IC}
\centering
\label{tab:coverage}
\begin{tabular}{l|l|l|l|l|}
\cline{2-5}
                                 & \multicolumn{2}{l|}{CHT}                                      & \multicolumn{2}{l|}{TARC}                                      \\ \hline
\multicolumn{1}{|l|}{Algorithms} & Cost & \begin{tabular}[c]{@{}l@{}}Delivery \\ ratio (\%)\end{tabular} & Cost & \begin{tabular}[c]{@{}l@{}}Delivery\\ ratio (\%)\end{tabular} \\ \hline
\multicolumn{1}{|l|}{MSC}  & 4    & 86.0512                                                        & 13   & 87.0903                                                       \\ \hline
\multicolumn{1}{|l|}{BC} & 25   & 84.4760                                                        & 459  & 87.8831                                                       \\ \hline
\multicolumn{1}{|l|}{IC}  & 18   & 85.9515                                                        & 237  & 88.4032                                                       \\ \hline
\end{tabular}
\end{table}

\subsection{Minimizing Latency}
We evaluate the effectiveness of the CELF-MDD algorithm (described in Section~\ref{subsec:celf_im}) at minimizing the overall message delay in the network model. 
We first compare the difference in run-time efficiency between Greedy-MDD and CELF-MDD. Figures~\ref{fig:lookups} and~\ref{fig:running_time} show the average number of lookups and running-times for a gateway budget of up to 15 when working with the CHT network. 
The number of lookups for each budget refers to the number of times the influence function needs to be computed before a gateway was selected at that stage. 
Greedy-MDD grows linearly because for each iteration, the influence gain for all gateways that have not been selected has to be computed. 
However for CELF-MDD, the number of lookup grows much slower due to it leveraging results from past computation as discussed in Section~\ref{subsec:celf_im}.

The performance of CELF-MDD was compared to betweenness centrality (BC) and in-degree centrality (IC) in terms of network latency minimization.
For each algorithm, the top $k$ stops generated were selected as edge nodes, where $k$ is the budget. 
The centrality metrics were computed on graph weighted with latency.
Simulations were run with the selected edge node set using the simulation parameters outlined in Table~\ref{sim_parameters}. 
Consistent with the influence function, the  upper bound value specified in Table~\ref{sim_parameters} is assigned as the penalty value (delay) for undelivered data.
Figure~\ref{budget_delay_cht} and~\ref{budget_delay_louisville} show the average network latency of various budgets using each algorithm. 

For CHT, we observe that CELF-MDD consistently outperforms both BC and IC by $\approx 20$ minutes or higher. 
In addition, there is very little decrease in delay after first 5 edge nodes have been selected when using CELF-MDD. 
For TARC, we observe that CELF-MDD consistently outperforms both BC and IC by $\approx 45$ minutes or higher. 
There is also minimal decrease in delay after the first 9 edge nodes have been selected when using CELF-MDD. 
For CHT and TARC, the CELF-MDD algorithm can effectively serve the whole network with 5 and 9 well placed edge nodes, respectively.  

\begin{table}[htbp]
\centering
\caption{Simulation and Scenerio Parameters}
\label{sim_parameters}
\begin{tabular}{|l|l|}
\hline
Random generator seeds & 0:1:100\\ \hline
Simulation start time & 1:00:00\\ \hline
Simulation end time & 24:00:00\\ \hline
Time penalty/maximum-latency (hours)  & 25\\ \hline
Number of sensors & $30\% \times  |stops|$ \\ \hline
Sensor data generation frequency (minutes) & $U(1, 120)$\\ \hline
Number of sensor scenarios & 5  \\ \hline
Number of sensors for scenarios & $U(30, 40)$ \\ \hline
\end{tabular}
\end{table}

\section{Conclusion} \label{sec:discussion}
This work addresses the problem of efficient edge node placement in low-cost smart cities that opportunistically utilize public transit networks as data mules. 
We introduced several optimization algorithms and compared them to traditional network centrality measures. 
Experiments were carried out using public transport networks in two cities in the United States; Chapel-Hill and Louisville. 
The results show that our algorithms outperform traditional centrality measures by reducing network latency and ensuring coverage at minimal cost, indicating that our algorithms are effective in determining the best locations to place edge nodes, minimize delivery delay and minimize cost in low-cost smart cities that opportunistically utilize public transit networks as data mules. 

\bibliographystyle{IEEEtran}
\bibliography{conference}

\end{document}